\documentstyle[preprint,prb,aps]{revtex}
\begin{document}
\draft
\title{Applying the linear $\delta$-expansion  to disordered systems}
\author{M. P. Blencowe\cite{miles} and A. P. Korte}
\address{The Blackett Laboratory, Imperial College of Science, Technology and 
Medicine,
London~SW7~2BZ, United Kingdom}
\maketitle
\begin{abstract}
We apply the linear $\delta$-expansion (LDE), originally developed as a 
nonperturbative, 
analytical approximation scheme
in quantum field theory, to problems involving noninteracting electrons in 
disordered solids. The initial idea that the LDE method might be applicable to 
disorder is suggested by 
the  resemblance of the supersymmetric field theory formalism for quantities
such as the disorder-averaged density of states and conductance to the  path 
integral expressions
for the $n$-point functions of $\lambda\phi^4$ field theory, where the  LDE has 
proved a 
successful method of approximation.  The field theories relevant 
for disorder have several unusual  features which have not been considered 
before, however, 
such as anticommuting fields with Faddeev-Popov (FP) rather than Dirac-type 
kinetic energy terms, 
imaginary couplings and  Minkowskian field coordinate metric.
Nevertheless we show that the  LDE method can be successfully generalized to 
such field systems.
As a first test of the method and also to give some understanding of its 
origins, we
calculate to third order in the LDE  the ground state energy of a supersymmetric 
anharmonic oscillator
with FP kinetic term
and  real anharmonic coupling strength of arbitrary magnitude. Strong evidence 
for 
the convergence of the LDE is obtained. 
We then calculate  to second order in the LDE 
the disorder-averaged
density of states of a one dimensional system  and find  
even at first order more accurate
results than the commonly used self-consistent Born approximation. 
In
the final part we outline one possible way in which the LDE method might be 
applied to the conductance, 
using  as supporting example a zero-dimensional  model  with  Minkowskian field 
coordinate metric.  
Further directions for
research are discussed in the conclusion.         
\end{abstract}


\section{Introduction}

Interacting quantum field systems rarely allow exact, closed-form expressions 
for
their observable quantities. In nearly every case some method of approximate 
solution must be
employed. The most common approximation  involves developing a series expansion 
of the  
quantity of interest in increasing powers of the coupling constant multiplying 
the 
interaction term in the field Lagrangian. However,  such a series is not 
expected to give a good 
approximation
 when the coupling constant in the appropriate
dimensionless units is not small. Even when the coupling is small, there are 
situations where
the series approximation does not work, such as when the quantity of interest is 
nonanalytic at zero
coupling. Given the relevance of strongly interacting quantum field systems for 
describing processes
occurring in nature, a worthwhile and stimulating challenge has been to find 
alternative, analytical
approximation
schemes which must necessarily be nonperturbative in the physical coupling 
constants of the field 
systems.

The linear $\delta$-expansion (LDE) is one such scheme which has been 
successfully  applied to
problems in, for example, $\lambda\phi^4$ theory\cite{stancu} and quantum 
chromodynamics.\cite{solov,akeyo} In outline, the LDE 
method replaces
the action $S$ of the  field theory with a modified action $S_{\delta}$ which 
interpolates linearly
between a soluble action $S_0$ depending on a variational parameter $\Omega$
and the original $S$, i.e.,
\begin{equation}
S_{\delta}=(1-\delta )S_0 +\delta S.
\label{Sdeltadef}
\end{equation}
The Green function of interest $G$ is evaluated as a
power series in the artificial parameter $\delta$ up to the desired order $N$ 
and then $\delta$ is
set equal to 1. Unlike the exact Green function $G$, the truncated Green 
function $G_N$ will
depend on $\Omega$. The latter is fixed at each order $N$
 by applying the principle of minimal sensitivity 
(PMS):\cite{stevenson}
\begin{equation}
\left.\frac{\partial G_N}{\partial\Omega}\right|_{\Omega_N}=0.
\label{PMS}
\end{equation}    
The PMS condition is crucial, in that it provides the
nonperturbative dependence on the physical coupling parameter and in several 
cases has been
shown to ensure convergence of the sequence of approximants $G_N(\Omega_N)$. 
There is considerable arbitrariness in the choice of $S_0$. Some choices will 
result in 
more rapid convergence of the sequence or simpler calculations to perform, 
whereas other choices may not
work at all.  At the very least, $S_0$ should have
some resemblance to the original action $S$.  
Proofs of the convergence of the LDE method have been obtained in several 
different ways for certain Green
functions of the quantum 
anharmonic oscillator.\cite{arvanitis,guida,pernice,janke} These proofs have 
also
given us insights into how the method works. For higher dimensional field 
theories no convergence proofs
have been constructed so far  and we must rely on comparison with numerical 
methods.

In this paper we apply the LDE method to a different problem:  the  quantum 
dynamics of 
a single
electron moving in a random background potential. 
The idea of using the LDE method  comes from the 
so-called 
supersymmetric field formulation of the problem,\cite{mckane,efetov} 
in which  quantities characterizing the 
quantum dynamics, such as the 
disorder-averaged
density of states and conductance, are represented as path integrals over both 
commuting and anticommuting
field coordinates of some interacting field system. For a random
potential which is Gaussian delta-function correlated 
the corresponding supersymmetric field actions resemble the usual 
$\lambda\phi^4$ action, hence
suggesting the
possibility of applying the LDE method to the former systems as well. Of course, 
there are already 
several well-established approximation methods for studying the quantum dynamics 
of an 
electron in a random potential. Nevertheless, we  thought it would be 
interesting to see whether  
a method originally developed for the study of relativistic quantum field 
systems 
could be successfully adapted to random electron systems. 
First indications are that the LDE method in fact performs rather
well in comparison with  established approximation methods.

The supersymmetric field actions have several unusual features which have  not 
been considered before in LDE investigations. 
The kinetic term for the anticommuting fields is of Faddeev-Popov (FP)
 rather than the usual Dirac form (i.e., $\partial_t \chi^* \partial_t \chi$ 
instead of 
$\chi^* \partial_t \chi$). 
Also, the interaction terms appearing in the actions are imaginary and, for the 
action 
associated with the
disorder-averaged conductance, we have a 
Minkowskian field  metric [i.e., $\eta_{m n}\Phi_m^{\dag}\Phi_n$, with 
$\eta_{m n}= (-1)^m\delta_{m n}$; $m,n=1,2$]. In the following
 sections we show how the LDE method generalizes
to such systems.

In Sec.\ II we  evaluate  to third order in $\delta$ the ground state energy of 
a supersymmetric
anharmonic oscillator with FP kinetic term and real, positive anharmonic 
coupling parameter. 
The action for this
system is in fact the same as that associated with the  one-dimensional density 
of states for
delta-function-correlated disorder, the only difference being that the coupling 
is real instead of 
imaginary. Although the ground state energy is of little relevance for the 
random electron dynamics, it
is the simplest quantity on which to test the LDE method. Furthermore, the 
calculation gives some 
 understanding of the method's origins by showing that to first order in 
$\delta$ it  is equivalent to 
choosing a Gaussian wavefunction
with variational parameters which are fixed by requiring that the energy 
expectation value
 be a minimum. As we shall see, the LDE method reveals a rather remarkable 
dependence
of the ground state 
energy  on the frequency and anharmonic coupling parameters, making the 
supersymmetric 
anharmonic oscillator an interesting system in its own right.

The LDE method is applied to the disorder-averaged energy density of states in 
Sec.\ III. 
We first consider a ``zero-dimensional"
density of states model (i.e.,  path integral replaced by ordinary integral), 
for which we can easily
go to high order in $\delta$ and obtain  support for the convergence of the 
expansion.
We then evaluate to 
second order in $\delta$ the one-dimensional density of states for 
delta-function-correlated disorder
and compare with the exact expression,
as well as with the self-consistent Born approximation (SCBA).  
While the use of the LDE was motivated by the supersymmetric field formulation, 
it is  in fact not necessary
to calculate the density of states within this formulation. Indeed, we show that 
the same series 
approximation
is obtained by expanding in $\delta$  the averaged one-electron Green function 
expressed directly in
terms of a $\delta$-modified single-electron Hamiltonian. Note that the LDE 
method is not restricted to low dimensions, nor to just 
delta-function-correlated random potentials. 

Because the supersymmetric action associated with
the disorder-averaged conductance has Minkowskian field coordinate metric, 
the action must be $\delta$-modified in a way which is quite different from that 
of the
density of states action. 
In Sec.\ IV, we describe one possible $\delta$-modification.
In contrast with the density of states, 
the supersymmetric field formulation is essential for the application of  
the LDE method to the conductance: there is no
modified physical Hamiltonian which yields the same series expansion in 
$\delta$. 
As a first test, we apply the method to a zero-dimensional conductance model.

In the conclusion we suggest further directions for research. This includes the 
application of the 
LDE method to the supersymmetric non-linear sigma model in order to determine 
the conductivity
critical exponent.
   
\section{The Supersymmetric Anharmonic Oscillator}

In this section we  use the LDE method to determine the ground state energy of a 
quantum, supersymmetric
anharmonic oscillator. The classical Lagrangian is
\begin{equation}
L=\frac{1}{2}\dot{\Phi}^{\dag}\dot{\Phi} - \frac{\omega^2}{2}\Phi^{\dag}\Phi 
-\lambda (\Phi^{\dag}\Phi)^2,
\label{susyaho}
\end{equation}
where    
\begin{equation}\Phi = \left( \begin{array}{c}\chi \\ s \end{array} \right)
\text{,\ \ } \Phi^{\dag} =\left(\chi^*\ s^*\right)
\label{supervectdef} 
\end{equation}
is a supervector coordinate with anticommuting component $\chi$ and commuting 
component $s$.
In all our calculations involving anticommuting variables we follow the rules 
and conventions of 
Efetov,\cite{efetov} the only difference occurring in the definition of complex 
conjugation for
anticommuting variables. Our 
definition is
\begin{equation}
(\chi_1 \chi_2)^* =\chi_2^* \chi_1^*\text{,\ \ } (\chi^* )^* =\chi.
\label{complexdef}
\end{equation}               
The above Lagrangian is motivated by the supersymmetric formulation of the 
disorder-averaged
 density of states (see Sec.\ III).
In one dimension and for   delta-function-correlated random potential, the 
supersymmetric 
Lagrangian associated with the density of states differs from Eq.\ 
(\ref{susyaho}) only in having an
imaginary instead of real coupling (as well as an unimportant overall sign and 
constant coefficients).
Bohr and Efetov\cite{bohr,efetov} also investigated the above 
supersymmetric Lagrangian and found a
closed form expression for an eigenstate with eigenvalue zero which they assumed 
to be the ground state. 
As we shall see, however,
this assumption is correct only for $\omega^2$ larger than a certain negative 
value.  

In order to  quantize  the oscillator we must first write down the
Hamiltonian. In terms of  real  coordinates $s_1$, $s_2$, $\chi_1$, and 
$\chi_2$, where $s=s_1+is_2$
and $\chi=\chi_1+i\chi_2$, the Hamiltonian is
\begin{eqnarray}
H & = & \dot{s}_i p_i + \dot{\chi}_i \pi_i  -L\nonumber\\
& = & \frac{1}{2}(p_1^2 +p_2^2) +\frac{1}{2}\omega^2(s_1^2 +s_2^2) -i \pi_1\pi_2 
+i\omega^2\chi_1\chi_2+
\lambda (2 i\chi_1\chi_2 +s_1^2 +s_2^2 )^2,
\label{hamiltonian}
\end{eqnarray}
where the momenta are defined as follows:
\begin{equation}
p_i =\frac{\partial L}{\partial\dot{s}_i}=\dot{s}_i\text{,\ \ }\pi_i =
\frac{\roarrow{\partial} L}{\partial\dot{\chi}_i}=i\epsilon_{ij}\dot{\chi}_j.
\label{momenta}
\end{equation}
The oscillator can now be straightforwardly quantized  using the correspondence 
principle, i.e., by 
associating with the position coordinates and conjugate momenta  operators 
acting on some state space
and satisfying canonical (anti)commutation relations:
\begin{equation}
[\hat{p}_i,\hat{s}_j]=-i\delta_{ij} \text{\ \ and \ \  
}\{\hat{\pi}_i,\hat{\chi}_j\}=-i\delta_{ij},
\label{commutators}
\end{equation}
with all other (anti)commutators vanishing and where we have  set $\hbar=1$. 
Note from 
Eq.\ (\ref{momenta})
that $\hat{\pi}_i$ is antihermitian, i.e., $\hat{\pi}^{\dag}_i =-\hat{\pi}_i$.

It will be convenient to work with two different state space bases. In one, the 
states are
wavefunctions depending on the position coordinates, 
$\psi(s_1,s_2,\chi_1,\chi_2)$, with scalar product
defined as
\begin{equation}
\langle\psi|\phi\rangle=-i\int d\chi_1 d\chi_2 ds_1 ds_2 \ 
\psi^*(s_1,s_2,\chi_1,\chi_2)
\phi(s_1,s_2,\chi_1,\chi_2),
\label{innerproduct}
\end{equation}
and the position and momentum operators represented as follows:
\begin{equation}
\hat{s}_i\leftrightarrow s_i\text{,\ \ }\hat{p}_i\leftrightarrow 
-i\frac{\partial}{\partial s_i}
\text{,\ \ }\hat{\chi}_i\leftrightarrow\chi_i\text{,\ \ 
}\hat{\pi}_i\leftrightarrow
-i\frac{\roarrow{\partial}}{\partial \chi_i}.
\label{posmomrep}
\end{equation}    

The energy eigenstates of the free  Hamiltonian ($\lambda=0$ and $\omega^2 >0$) 
form a second useful 
basis. This basis is most easily constructed using creation/annihilation 
operators, defined in terms 
of the position and momentum operators as follows:
\begin{eqnarray}
s_1 & = & \frac{1}{\sqrt{2\omega}}(a+a^{\dag})\nonumber\\
p_1 & = & -i\sqrt{\frac{\omega}{2}} (a-a^{\dag})\nonumber\\   
s_2 & = & \frac{1}{\sqrt{2\omega}}(b+b^{\dag})\nonumber\\
p_2 & = & -i\sqrt{\frac{\omega}{2}} (b-b^{\dag})\nonumber\\   
\chi_1 & = & \frac{1}{2\sqrt{\omega}}(c+c^{\dag}+d+d^{\dag})\nonumber\\
\pi_1 & = & -i\frac{\sqrt{\omega}}{2}(c+c^{\dag}-d-d^{\dag})\nonumber\\ 
\chi_2 & = & -i\frac{1}{2\sqrt{\omega}}(c-c^{\dag}-d+d^{\dag})\nonumber\\
\pi_2 & = & -\frac{\sqrt{\omega}}{2}(c-c^{\dag}+d-d^{\dag}),
\label{creatannihdef}
\end{eqnarray}
where we have omitted the hats on the operators. From these definitions and the 
canonical 
(anti)commutation relations (\ref{commutators}), we have
\begin{equation}
[a,a^{\dag}]=1\text{,\ \ }[b,b^{\dag}]=1\text{,\ \ }\{c,c^{\dag}\}=1\text{,\ \ 
}\{d,d^{\dag}\}=-1,
\label{creatannihcommut}
\end{equation}
with all other (anti)commutators vanishing.
Using Eqs.\ (\ref{creatannihdef}) and (\ref{creatannihcommut}) to express the 
noninteracting
part of the Hamiltonian in terms of the creation/annihilation operators, we find 
that
\begin{equation}
H_0 =\omega (a^{\dag}a +b^{\dag}b +c^{\dag}c -d^{\dag}d).
\label{freehamiltonian}
\end{equation}
The energy eigenstates are obtained in the usual way. We first introduce a state 
$|0\rangle$,
 satisfying
\begin{equation}
a|0\rangle=b|0\rangle=c|0\rangle=d|0\rangle=0
\label{groundstate}
\end{equation}
and $\langle 0|0\rangle=1$.
>From Eqs.\ (\ref{freehamiltonian}) and (\ref{groundstate}) we see that 
$|0\rangle$ is an energy
eigenstate with eigenvalue 0. All other eigenstates are obtained by acting on 
$|0\rangle$ with
the creation operators $a^{\dag}$, $b^{\dag}$, $c^{\dag}$, and $d^{\dag}$. The 
eigenstate
\begin{equation}
|m,n,j,k\rangle = \frac{(a^{\dag})^m (b^{\dag})^n (c^{\dag})^j (d^{\dag})^k}
{\sqrt{m!n!}}|0\rangle
\label{eigenstate}
\end{equation}
has eigenvalue
\begin{equation}
E_{mnjk}=(m+n+j+k) \omega \geq 0,
\label{eigenvalue}
\end{equation}
for  integers $m$, $n$, $j$, and $k\geq 0$,  
so that $|0\rangle$  is in fact the ground state of the free oscillator.
Note that, since $c^{\dag}$ and $d^{\dag}$ anticommute,  $j$ and $k$ are either 
zero or one.
Note also that, because of the minus sign entering in the anticommutation 
relation 
(\ref{creatannihcommut}) for $d$ and $d^{\dag}$, eigenstates with $k=1$ have 
negative norm. The occurrence
of negative norm states in the state space can be traced back to having an
anticommuting coordinate kinetic term with two time derivatives, 
$\dot{\chi}^*\dot{\chi}$,
in Eq.\ (\ref{susyaho}). The  FP ghost fields, which occur in the quantization 
of Yang-Mills fields,
have similar properties.\cite{kugo} Negative norm  states can be avoided by 
using instead a Dirac 
kinetic term, $\chi^*\dot{\chi}$. However, since the FP kinetic term is the 
relevant one 
for disorder and the  Lagrangian (\ref{susyaho}) is  not meant
to describe a physical oscillator, we shall live with the negative norm states.

Using Eqs.\ (\ref{groundstate}), (\ref{creatannihdef}), and
(\ref{posmomrep}) we
find that
the ground state  for the free Hamiltonian has the following form in the
position coordinate basis:
\begin{equation}
\psi_0(s_1,s_2,\chi_1,\chi_2)=\frac{1}{\sqrt{2\pi}}\exp\left(-\frac{1}{2}\omega\
Phi^{\dag}\Phi
\right)=\frac{1}{\sqrt{2\pi}}\exp\left[-\frac{1}{2}\omega\left(s_1^2 +s_2^2 +
2i\chi_1\chi_2\right)\right].
\label{coordgroundstate}
\end{equation}
This wavefunction is just the supersymmetric generalization of the Gaussian
function.

Let us now return to the problem of determining the ground state energy of
the  supersymmetric oscillator Hamiltonian for $\lambda >0$. In the
case of the ordinary,
commuting coordinate anharmonic oscillator it is well known that the minimum
energy expectation value
for  a Gaussian wavefunction with variable frequency and coordinate shift
parameters gives a good
approximation to the ground
energy.\cite{stevenson2}
A natural choice for the supersymmetric oscillator trial wavefunction is then
\begin{equation}
\psi(s_1,s_2,\chi_1,\chi_2)=\frac{1}{\sqrt{2\pi}}\left(\frac{
\Omega_1\Omega_2}{\nu^2}\right)^{1/4}\exp\left[-\frac{1}{2}\Omega_1
(s_1-s_0)^2-\frac{1}{2}\Omega_2 s_2^2 -i\nu\chi_1\chi_2\right],
\label{susytrialfunction}
\end{equation}
where $s_0$ is the shift parameter and $\Omega_1$, $\Omega_2$, and $\nu$
are the  frequency parameters.
A nonzero shift parameter is necessary in order to  approximate the ground
energy for $\omega^2 <0$ and $|\omega^2|$ large, since in this case  the 
commuting coordinate
part of the potential in (\ref{hamiltonian}) has the ``Mexican hat" form
and the  ground-state wavefunction peaks at  the rim of  the hat,
instead of at the center. The expectation value $\langle\psi|H|\psi\rangle$
is most easily worked out  in the position coordinate representation
using Eqs.\ (\ref{hamiltonian}), (\ref{posmomrep}),
(\ref{susytrialfunction}), and (\ref{innerproduct}). The minimum
expectation value with respect to variations in the parameters $s_0$,
$\Omega_1$, $\Omega_2$, and $\nu$ then gives an approximation to the ground
state energy.

The above method  provides only a one-off approximation, however. A
considerable improvement would be  a  scheme in which the expectation value
$\langle\psi|H|\psi\rangle$, for $|\psi\rangle$ given by Eq.\
(\ref{susytrialfunction}), appears as the first-order term in a perturbation
series.  As shown by Stevenson for  the ordinary anharmonic oscillator  (see 
Sect. V of Ref.\
\onlinecite{stevenson2}),
such a scheme can in fact be realized.  Generalizing  to the supersymmetric
oscillator, the scheme involves first modifying Hamiltonian
(\ref{hamiltonian}) as follows:
\begin{eqnarray}
H_{\delta}&=&\frac{1}{2} (p_1^2 +p_2^2 ) +\frac{1}{2}\Omega_1^2 s_1^2 +
\frac{1}{2}\Omega_2^2 s_2^2 -i\pi_1\pi_2 +i\nu^2\chi_1\chi_2
+\frac{1}{2}\omega^2 s_0^2 +\lambda s_0^4\nonumber\\
& &+\delta\left[\frac{1}{2}\omega^2 (s_1 +s_0)^2 +\frac{1}{2}\omega^2
s_2^2 +i\omega^2\chi_1\chi_2
+\lambda\left(2i\chi_1\chi_2 +(s_1 +s_0)^2 +s_2^2\right)^2\right.\nonumber\\
& &-\left.\frac{1}{2}\omega^2 s_0^2
-\lambda s_0^4 -\frac{1}{2}\Omega_1^2 s_1^2 -\frac{1}{2}\Omega_2^2 s_2^2
-i\nu^2\chi_1\chi_2\right],
\label{hamiltoniandelta}
\end{eqnarray}
and then solving for the lowest energy eigenvalue of this modified
Hamiltonian using the usual Rayleigh-Schr\"{o}dinger (RS)   perturbation
procedure with $\delta$ serving as expansion parameter.
Note that $H_{\delta}$ interpolates linearly between a harmonic oscillator
Hamiltonian for $\delta=0$ and the original anharmonic Hamiltonian
(\ref{hamiltonian})  with coordinate redefinition $s_1 \rightarrow s_1
+s_0$   for $\delta=1$. Hence the name: ``Linear Delta Expansion". To first
order in $\delta$, the RS series approximation to the ground energy is (with 
$\delta$ set equal to 1):
\begin{equation}
E_0\approx E_0^{(0)} +\langle 0|H_{\rm int}|0\rangle,
\label{groundapprox}
\end{equation}
where $E_0^{(0)}$ and $|0\rangle$ are the ground energy eigenvalue and
eigenstate, respectively, of the harmonic oscillator Hamiltonian
$H_{\delta=0}$ and $H_{\rm int}=H_{\delta=1}-H_{\delta=0}$. The most
straightforward way to work out Eq.\ (\ref{groundapprox}) is to first
express $H_{\delta}$ in terms of creation/annihilation operators using
relations (\ref{creatannihdef}) with the appropriate frequency changes.
After some calculation we find that Eq.\ (\ref{groundapprox}) indeed
coincides with the expectation value $\langle\psi|H|\psi\rangle$. However,
we now have a systematic procedure which allows us to go beyond the
Gaussian variational approximation:  $E_0$ for the Hamiltonian $H_{\delta}$
is evaluated up to the desired order $N$ in $\delta$ using the RS
perturbation method and the frequency and shift  parameters are then fixed by
minimizing the order $N$ approximation to $E_0$. The calculations proceed
in much the same way as for the ordinary anharmonic oscillator, the main
difference arising from the negative norm eigenstates in the
state sums  which appear at second  order and higher.  The negative norm
states are best dealt with by making explicit the eigenstate normalization
factors  in the RS series expansion formula.

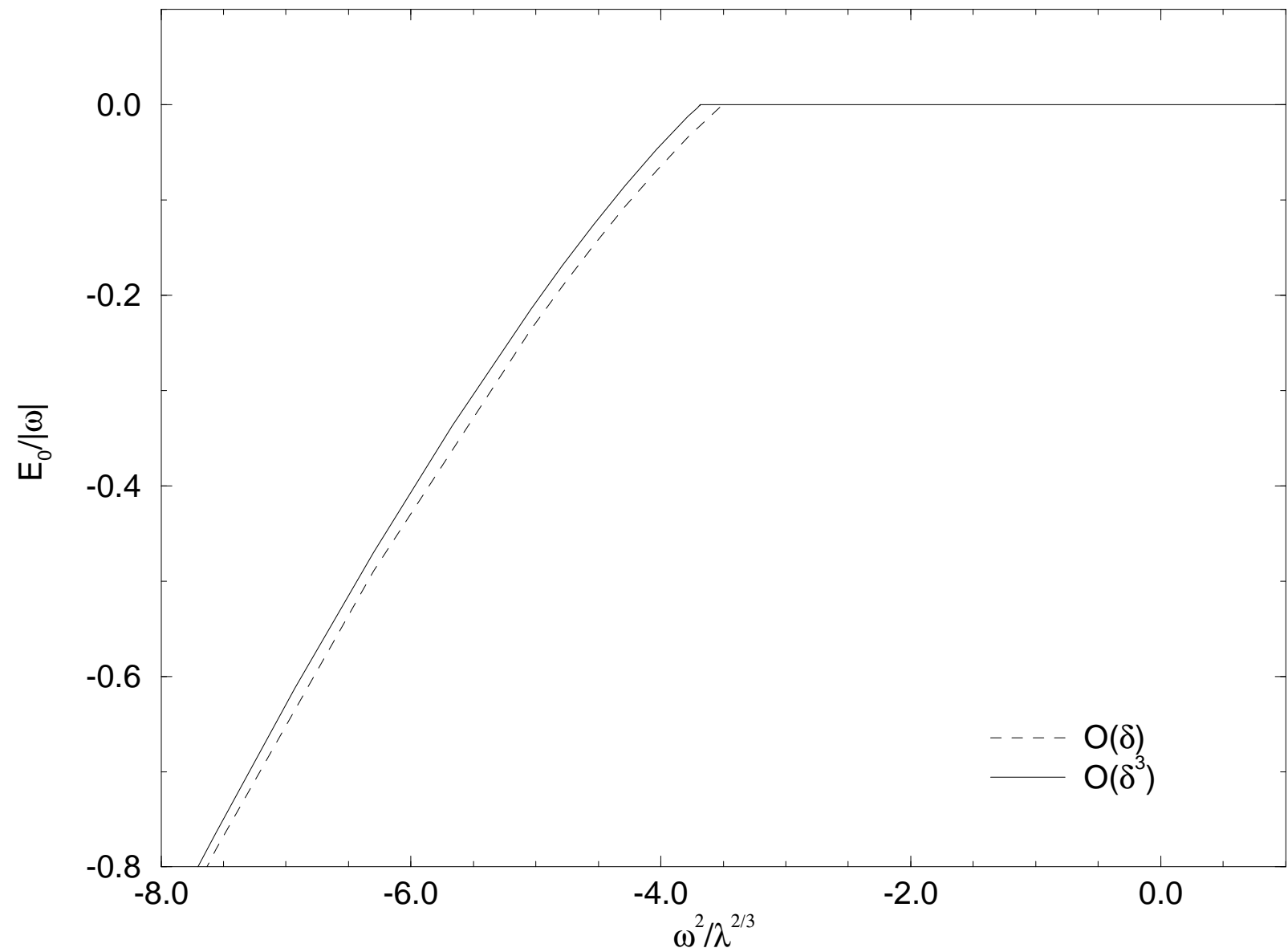
\begin{figure}
\caption{Ground state energy in units of $|\omega|$ versus
$\omega^2/\lambda^{2/3}$. The dashed line is the order $\delta$ approximation 
and the solid 
line the order $\delta^3$ approximation.}
\label{fig1}
\end{figure}

In Fig.\ \ref{fig1} we plot the results of the order $\delta$ and
$\delta^3$ ground state energy calculations. [To order $\delta^2$, there
were no parameter values for which $E_0$ was stationary. This is a common
occurrence in LDE calculations, where  stationary points may  exist  for odd 
(even)
orders only.] The analogous results for the ordinary anharmonic oscillator
are given in Ref.\ \onlinecite{stevenson2}. For the range
$\omega^2/\lambda^{2/3}\gtrsim -3.69$, the ground energy is exactly zero to
order $\delta^3$. The
frequency  parameters satisfy $\Omega_1=\Omega_2=\nu$, while the shift
parameter $s_0=0$. For $\omega^2/\lambda^{2/3}\lesssim -3.69$, $E_0$ is
stationary for a choice of parameters with the same pattern as above, again
giving $E_0=0$. However,  there is another choice of parameters having the
pattern $\Omega_1\neq\Omega_2=\nu$ and $s_0\neq 0$ for which $E_0$ is also
stationary but less than zero, hence providing a closer approximation to
the ground state energy.

An independent check of the results for $\omega^2<0$ and $|\omega^2|$ large
is obtained by expressing the Hamiltonian (\ref{hamiltonian}) in polar
coordinates and expanding the commuting coordinate potential to quadratic
order in the radial coordinate difference $r-r_0$, where $r_0$ is the
location of the potential minimum. To leading order, the ground state
energy of the resulting harmonic oscillator Hamiltonian  is
\begin{equation}
E_0\approx -\frac{\omega^4}{16\lambda}.
\label{groundasympt}
\end{equation}
We have verified that the order $\delta$ and $\delta^3$  approximations
indeed tend to (\ref{groundasympt}) as $\omega^2\rightarrow -\infty$. While not 
constituting a proof,
this result taken together with the closeness of the order $\delta$ and 
$\delta^3$ approximations
strongly suggests that the $\delta$-expansion converges.

 The existence of a region where the ground energy is exactly zero is
reminiscent of
quantum mechanical systems with time translation supersymmetry [i.e., the
square of the supersymmetry operator equals the Hamiltonian  (for a review,
see e.g., Ref.\ \onlinecite{genden})]. For such systems, invariance of the
ground state under  supersymmetry transformations implies the vanishing
of the ground state energy. On the other hand, if the ground state is
noninvariant, time translation supersymmetry is spontaneously broken and we
have  a nonvanishing  (positive) ground state energy. It is tempting,
therefore, to speculate that  time translation supersymmetry is broken for
$\omega^2$ below $-3.69 \lambda^{2/3}$ and unbroken  above this value.
However, the  Hamiltonian (\ref{hamiltonian}) is  known only to have rotational
supersymmetry (i.e., the square of the supersymmetry operator equals the
angular momentum operator corresponding to rotations in the $s_1$--$s_2$
coordinate plane) and no  conclusions can be drawn concerning the ground
energy from the  invariance properties of the ground state under rotational
supersymmetry. It would be of interest to try to understand the reasons for  the
 rather remarkable dependence of the ground state energy on
$\omega^2$ shown in Fig.\ \ref{fig1}.

\section{The density of states}

Consider an electron in a $d$-dimensional random potential, described by the 
Hamiltonian
\begin{equation}
H=\frac{p_i p_i}{2m} +V({\bf r}) -E,\ \ i=1, \ldots, d, 
\label{physicalhamiltonian}  
\end{equation}
where $E$ is the Fermi energy and the potential $V$ is Gaussian-distributed  
with delta-function 
correlation:
\begin{equation}
\overline{V({\bf r})}=0,\ \ \overline{V({\bf r}) V({\bf r}')}=\lambda 
\delta({\bf r}-{\bf r}').
\label{correlation}
\end{equation}
The overline denotes disorder averaging and the parameter $\lambda$ 
characterizes the strength of the
disorder. The supersymmetric formulation of the averaged energy density of 
states per unit volume for this
system is\cite{efetov}
\begin{equation}
\overline{\rho(E)}=-\frac{1}{\pi}{\rm Im}\ \overline{G^{(+)}({\bf r},{\bf 
r};E)}=
\frac{1}{\pi}{\rm Re}\int D\Phi^{\dag}D\Phi\ s({\bf r})s^*({\bf r})\exp(i S),
\label{density}
\end{equation}
where $G^{(+)}$ is the retarded one-electron Green function.
The supervector coordinate $\Phi$ is defined in Eq.\ (\ref{supervectdef}) and 
the supersymmetric
action $S$ is  defined as follows:
\begin{equation}
S=-\int d{\bf r}\left[\frac{\hbar^2}{2 m}\partial_i \Phi^{\dag}\partial_i \Phi -
(E+i\epsilon )\Phi^{\dag}\Phi 
-\frac{i\lambda}{2}\left(\Phi^{\dag}\Phi\right)^2\right].
\label{action}
\end{equation} 
Given the resemblance of this action to the ordinary $\lambda\phi^4$ action and 
the effectiveness of the LDE approximation method for studying various quantum 
properties of  
$\lambda\phi^4$ theory,\cite{stancu} it is natural to apply the LDE method to 
the density
of states of system (\ref{physicalhamiltonian}) as well. Of course, as we have 
already mentioned,
the supersymmetric action has several significant additional features. 
In the preceding section we showed how the LDE method could successfully 
accommodate some of 
these features in ground state energy calculations. 
In the following, we  further extend the LDE method, using it to approximate the
expression (\ref{density})  which is essentially the two-point function
of a supersymmetric $\lambda\phi^4$ system with imaginary coupling.

The first step in the LDE procedure is to ``$\delta$-modify" the action 
(\ref{action}). 
Recall that, for
the parameters satisfying $\Omega_1 =\Omega_2 =\nu$
and $s_0 =0$, the $\delta$-modified Hamiltonian (\ref{hamiltoniandelta}) gave 
the  
energy $E_0 =0$. Furthermore, Bohr and Efetov\cite{bohr} 
(see also Sect.\ 6 of Ref.\ \onlinecite{efetov})
showed, by expressing the $d=1$ density of states in terms of the eigenstates of 
Hamiltonian 
(\ref{hamiltonian}),  that  only the  eigenstate  with eigenvalue zero was 
relevant. 
Noting the form of the $\delta$-modified Hamiltonian (\ref{hamiltoniandelta})
for $\Omega_1 =\Omega_2 =\nu=\Omega$ and $s_0 =0$, we are led to consider the 
following  
$\delta$-modified action:
\begin{equation}
S_{\delta}=-\int d{\bf r}\left\{\frac{\hbar^2}{2 m}\partial_i 
\Phi^{\dag}\partial_i \Phi -
(\Omega+i\epsilon )\Phi^{\dag}\Phi +
\delta\left[ (\Omega-E)\Phi^{\dag}\Phi 
-\frac{i\lambda}{2}\left(\Phi^{\dag}\Phi\right)^2\right]\right\}.
\label{actiondelta}
\end{equation}
The original action (\ref{action})  is replaced by the modified action 
(\ref{actiondelta}) in 
Eq.\ (\ref{density}) and the series expansion in $\delta$ obtained. The final 
step is the application
of the PMS condition (\ref{PMS}) in order to fix the frequency parameter 
$\Omega$. Note that the
PMS condition is applied before taking the real part in expression 
(\ref{density}).

Again it is not necessary to implement the LDE method 
within the supersymmetric formulation of the density of states. 
Comparing the actions $S$ and $S_{\delta}$ and 
examining also  the correspondence between the action $S$ and the physical 
Hamiltonian 
(\ref{physicalhamiltonian}) with 
correlation relation (\ref{correlation}), 
we can immediately write down the modified physical Hamiltonian corresponding to 
the  action
$S_{\delta}$:   
\begin{equation}
H_{\delta}=\frac{p_i p_i}{2 m}  -\Omega + \delta\ (\Omega-E) +\delta^{1/2} 
V({\bf r}).
\label{1phdelta}
\end{equation}
Since the averaged potential satisfies $\overline{V({\bf r})}=0$, only integer 
powers in $\delta$
appear in the expansion.
If we then replace the physical Hamiltonian  by this modified Hamiltonian  in 
the definition
of the retarded one-electron Green function:
\begin{equation}
G^{(+)}_{\delta}({\bf r},{\bf r}')=
\langle{\bf r}|\left(-H_{\delta}
+i\epsilon\right)^{-1}|{\bf r'}\rangle,
\label{greendef}
\end{equation}
we recover the same series  expansion in $\delta$ as for the above 
supersymmetric formulation.
Although  the supersymmetric formulation is not required in order to apply the 
LDE approximation 
method to the density of 
states, the  formulation is of considerable value in suggesting the LDE method. 
Without 
the supersymmetric formulation, it is not obvious that modifying the physical 
Hamiltonian 
(\ref{physicalhamiltonian}) as
in Eq.\ (\ref{1phdelta}) is a useful step. Casting a quantity in a
different form can often suggest  new methods of approximation.

We first  consider a zero-dimensional density of states model which
provides  support for the convergence of the LDE method.
The model is essentially the dimensional reduction of expression 
(\ref{density}):
\begin{equation}
I=\int d\chi^* d\chi d s^* d s\ s^* s \exp i\left[-\alpha\Phi^{\dag}\Phi 
-\frac{\lambda}{2}\left(
\Phi^{\dag}\Phi\right)^2\right].
\label{zerodmodel}
\end{equation} 
Apart from the $s^* s$ factor outside the exponential, this integral is just the 
supersymmetric
version of the integral studied in Ref.\ \onlinecite{buckley}. 
Carrying out the integrations, we obtain
\begin{equation}
I=\sqrt{2\pi^3/\lambda}\ e^{\alpha^2/2\lambda}\ {\rm 
erfc}(\alpha/\sqrt{2\lambda}).
\label{zerodsolution}
\end{equation}    
(Note that, without the $s^* s$ factor, the integral would equal one.) Let us 
now approximate the
integral (\ref{zerodmodel}) using the LDE method. Modifying the argument of the 
exponential
as in Eq. (\ref{actiondelta}), we have
\begin{equation}
I_{\delta}=\int d\chi^* d\chi d s^* d s\ s^* s 
\exp i\left\{-\Omega\Phi^{\dag}\Phi +\delta\left[(\Omega-\alpha)\Phi^{\dag}\Phi 
-\frac{\lambda}{2}
\left(\Phi^{\dag}\Phi\right)^2\right]\right\}.
\label{deltazerodmodel}
\end{equation}       
The terms in the $\delta$-expansion are simple enough that a general expression 
can be written down
for the expansion up to arbitrary order $N$:
\begin{equation}
I_N =2\pi \sum_{n=0}^N \sum_{k=0}^n \sum_{j=0}^{2k}
(-1)^{k-j} 2^{-n} \lambda^{j-n}\Omega^{-j-1}(\Omega-\alpha)^{2n-j}
\frac{(2k)!}{k!(n-k)!(2k-j)!}.
\label{deltaexpand}
\end{equation}
We were able to evaluate $I_N$, with $\Omega$ fixed by the PMS condition, up to 
high order in
$N$ for a range of complex parameter values $\alpha$ and $\lambda$ and in each 
case found that
the sequence $I_N (\Omega_N)$ converged to the exact solution 
(\ref{zerodsolution}).

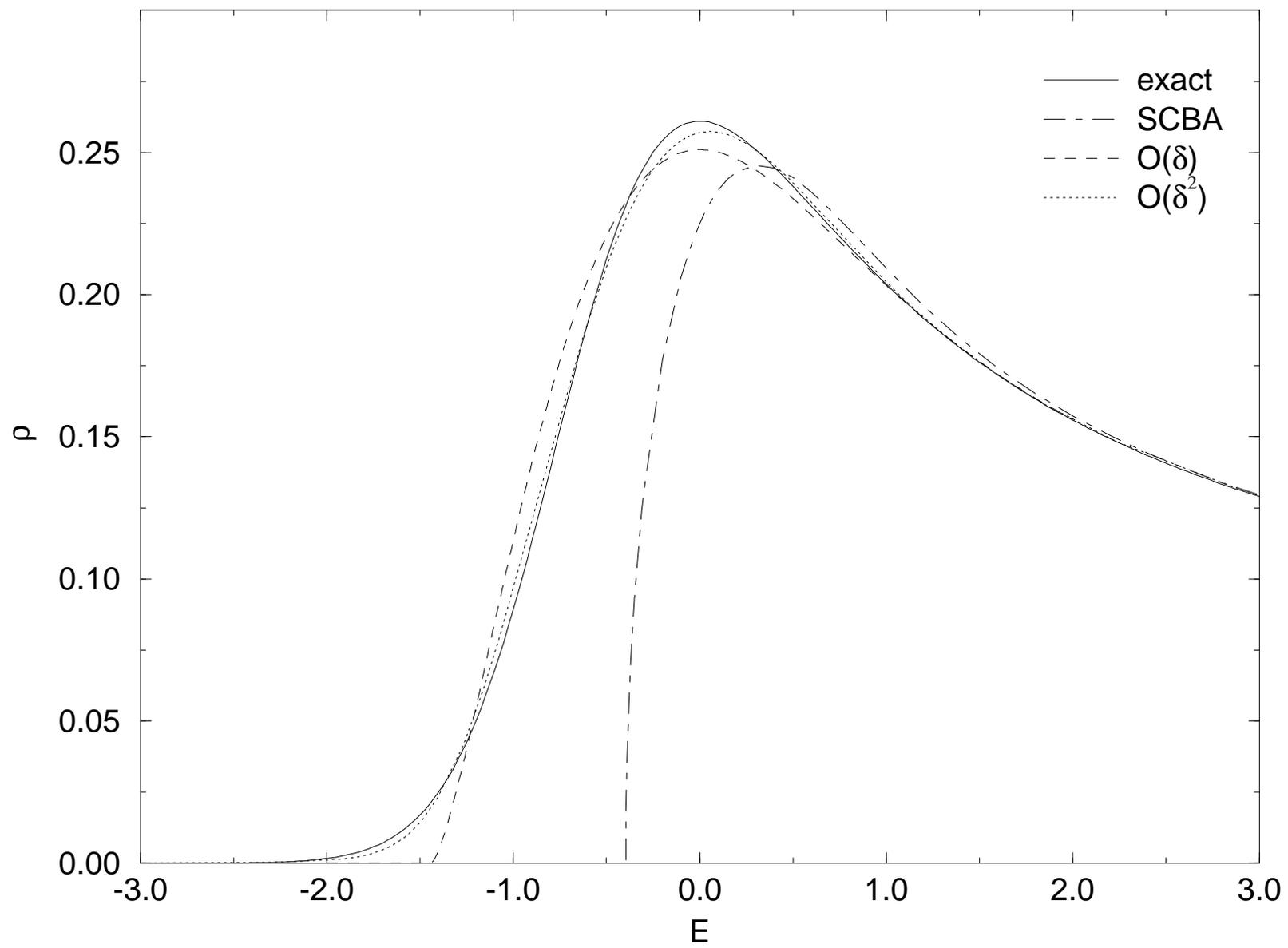
\begin{figure}
\caption{Energy density of states per unit length versus Fermi energy for
 disorder strength $\lambda=1/\protect\sqrt{2}$ in units $\hbar=m=1$. 
The solid line is the exact curve,
the dashed line the order $\delta$ approximation, the dotted line the order 
$\delta^2$ approximation,
and the dot-dashed line the self-consistent Born approximation.}
\label{fig2}
\end{figure}

Thus encouraged we use the LDE to approximate the $d=1$ density of states. This 
example has been 
solved exactly using a
variety of methods\cite{lloyd,halperin,thouless,bohr}  enabling an immediate 
check of the 
accuracy of the LDE. 
Expanding to first order in $\delta$ either  Eq.\ (\ref{density}) with $S$ 
replaced by 
$S_{\delta}$, or Eq.\ (\ref{greendef}) after  disorder averaging, we obtain 
(with 
$\delta$ set equal to 1):
\begin{equation}
\overline{\rho}_1= \frac{1}{2\pi\hbar}\sqrt{\frac{m}{2}}
{\rm Re}\left(3\Omega^{-1/2}-E 
\Omega^{-3/2}+i\sqrt{\frac{m}{2\hbar^2}}\lambda\Omega^{-2}\right).
\label{deltadensity}
\end{equation}
Setting to zero the derivative  with respect to $\Omega$  of the  term in 
brackets, 
we obtain the following PMS condition:
\begin{equation}
\Omega^{3/2}-E\Omega^{1/2}+\frac{2i}{3}\sqrt{\frac{2m}{\hbar^2}}\lambda=0.
\label{deltaPMS}
\end{equation}
This equation admits three independent solutions for $\Omega$. However, only one 
of these solutions
is physical, i.e., yields a positive nonzero density of states when substituted 
into 
Eq.\ (\ref{deltadensity}). To second order in $\delta$, we have
\begin{eqnarray}
\overline{\rho}_2=\frac{1}{8\pi\hbar}\sqrt{\frac{m}{2}}&{\rm 
Re}&\left(15\Omega^{-1/2}-5 E \Omega^{-3/2} 
+3 E^2 \Omega^{-5/2} +6 i \sqrt{\frac{2 m}{\hbar^2}}\right. \lambda 
\Omega^{-2}\nonumber
\\ & &\left.-4 i\sqrt{\frac{2m}{\hbar^2}}
\lambda E \Omega^{-3} -\frac{25 m}{8\hbar^2}\lambda^2 \Omega^{-7/2}\right).
\label{delta2PMS}
\end{eqnarray}
The PMS condition now admits six    independent solutions for $\Omega$ of which 
only one is  physical.
In Fig.\ \ref{fig2} we plot the order $\delta$ and $\delta^2$ approximations
to the $d=1$ density 
of states. Also shown  are the exact curve and the self-consistent Born 
approximation 
(SCBA) (see, e.g., Ref.\ \onlinecite{blencowe} and references therein). A 
comparison between 
the exact curve and the 
order $\delta$ and $\delta^2$ approximations provides strong evidence for the 
rapid  convergence 
of the LDE method. Note that even the lowest, order $\delta$, approximation is 
superior to the SCBA
result. The convergence appears to be slowest in the region of the exponentially 
decaying 
exact density of states tail. It is possible  that a different  
$\delta$-modification procedure 
from that used above can be found which gives better convergence in this region.
 
The analogous calculations for $d\geq 2$ should involve little extra difficulty. 
The only new feature is
the divergent nature of the terms in the $\delta$-expansion for $d\geq 2$, so 
that the procedures of
regularization and renormalization are required. Other 
investigations\cite{stancu}  
suggest that these procedures
are straightforward to implement within the LDE method. It should also be 
possible to use the LDE method
to approximate the disorder-averaged density of states for random potentials 
satisfying other 
correlation relations.

\section{The conductance}

In this section we present some  ideas concerning the application of the LDE 
method
to the disorder-averaged conductance. Consider a channel with cross-section $A$ 
and length $L$ which is
connected adiabatically at both ends to  reservoirs. The single electron 
Hamiltonian is given 
by Eqs.\ (\ref{physicalhamiltonian}) and 
(\ref{correlation}) with an additional hard-wall confining potential restricting 
the electron
to move within the cross-section $A$.
The supersymmetric formulation of the zero-temperature, disorder-averaged 
conductance is
\begin{eqnarray}
\overline{C}&=&\frac{e^2\hbar^3}{8\pi m^2 L^2}\int_{-L/2}^{+L/2}dx\int_A d{\bf 
r}_{\perp}
\int_{-L/2}^{+L/2}dx'\int_A d{\bf r}'_{\perp}\int D\Phi_1^{\dag} D\Phi_1 
D\Phi_2^{\dag} D\Phi_2
\nonumber\\
& & \left[s_1({\bf r})\partial_{x'}s_1^* ({\bf r}')s_2({\bf 
r}')\partial_{x}s_2^* ({\bf r})
+\partial_{x}s_1({\bf r})s_1^* ({\bf r}')\partial_{x'}s_2({\bf r}')s_2^* ({\bf 
r})\right.\nonumber\\
& &\left.-\partial_{x}s_1({\bf r})\partial_{x'}s_1^* ({\bf r}')s_2({\bf 
r}')s_2^* ({\bf r})
-s_1({\bf r})s_1^* ({\bf r}')\partial_{x'}s_2({\bf r}')\partial_{x}s_2^* ({\bf 
r})\right]\nonumber\\
& &\times\left(e^{iS^{+-}}+e^{iS^{-+}}+e^{iS^{++}}+e^{iS^{--}}\right),
\label{conductance}
\end{eqnarray}
where ${\bf r}= (x, {\bf r}_{\perp})= (x,y,z)$, and the actions $S^{ab}$ are 
defined as follows:
\begin{eqnarray}
S^{ab}=-\int dx 
\int_A d{\bf r}_{\perp}&&\left\{\eta_{mn}^{ab}\left[\frac{\hbar^2}{2m}\partial_i 
\Phi_m^{\dag}
\partial_i\Phi_n
-E \Phi_m^{\dag} \Phi_n \right]\right.\nonumber\\
&& \left. -i\epsilon\Phi_m^{\dag} \Phi_m -\frac{i\lambda}{2}
\left(\eta_{mn}^{ab}\Phi_m^{\dag} \Phi_n \right)
\left(\eta_{uv}^{ab}\Phi_u^{\dag} \Phi_v \right)\right\},
\label{condaction}
\end{eqnarray}
with the supervector fields $\Phi_m$, $m=1,2$, vanishing at the 
cross-section boundaries.
The field metrics are defined as follows:
\begin{equation}
\eta_{mn}^{++}=\delta_{mn},\  \eta_{mn}^{--}=-\delta_{mn},
\  \eta_{mn}^{+-}=(-1)^{m+1}\delta_{mn}, 
\  \eta_{mn}^{-+}=(-1)^m\delta_{mn};\ m,n=1,2.
\label{etadef}
\end{equation}
 Eq. (\ref{conductance}) follows from the formula for the disorder-averaged
conductance of a finite-length wire in terms of the averaged products 
of advanced and retarded Green functions (see, e.g., Ref.\ 
\onlinecite{fisher}) and the formulation of these products in terms of 
supersymmetric
path integrals.\cite{efetov} 

We can decompose Eq.\ (\ref{conductance}) into a sum of four terms, each 
involving a different action.
It is most natural to apply the LDE method to each of these four terms 
separately.  The 
action $S^{++}$ is very similar to the density of states action (\ref{action}), 
the only difference 
being the number of supervector fields. We therefore expect that the action 
$S^{++}$ 
can be $\delta$-modified with a single parameter $\Omega$ exactly as
in Eq.\ (\ref{actiondelta}). The series expansion in $\delta$ is obtained and 
the PMS condition 
applied  after all the integrals have been 
carried out.  The complex conjugate of the resulting LDE approximation
yields the  approximation to the $S^{--}$ term. 

The application of the LDE method to 
the terms involving the actions $S^{+-}$ and $S^{-+}$ is less straightforward as 
a consequence
of the Minkowskian field metric appearing in the actions. As a first step, we 
consider a zero-dimensional  
conductance model analogous to the density of states model defined in Eq.\ 
(\ref{zerodmodel}). These
models are particularly useful for testing the convergence of a given 
$\delta$-modification. 
Dimensionally reducing the $S^{+-}$ term in Eq.\ (\ref{conductance}), 
we arrive at the following model:
\begin{equation}
I=\int d\chi_1^* d\chi_1 d\chi_2^* d\chi_2 ds_1^* ds_1 ds_2^* ds_2\ 
s_1 s_1^* s_2 s_2^*\ \exp(iS),
\label{zerodcond}
\end{equation}
where
\begin{equation}
S=a\left(\Phi_1^{\dag}\Phi_1 -\Phi_2^{\dag}\Phi_2\right) +ib
\left(\Phi_1^{\dag}\Phi_1 +\Phi_2^{\dag}\Phi_2\right) +\frac{i\lambda}{2}
\left(\Phi_1^{\dag}\Phi_1 -\Phi_2^{\dag}\Phi_2\right)^2.
\label{zerodaction}
\end{equation}
The parameters $a$ and $b$ are real with $b>0$. We choose $b$ to be much smaller
than $\lambda$.   The term $ib$ is the analogue of the $i\epsilon$
term in Eq.\ (\ref{condaction}). In order that Eq.\ (\ref{zerodcond}) not 
diverge, however,
$b$ must be positive and non-zero, rather than infinitesimal. 
The integrals are readily carried out and we obtain 
\begin{equation}
I=\sqrt{2\pi^5/\lambda}\  b^{-1}\left\{e^{(b+ia)^2/2\lambda}\ {\rm 
erfc}[(b+ia)/\sqrt{2\lambda}]
+ e^{(b-ia)^2/2\lambda}\ {\rm erfc}[(b-ia)/\sqrt{2\lambda}]\right\}.
\label{zerodcondsolution}
\end{equation}
Note that Eq.\ (\ref{zerodcondsolution}) is just twice the real part of  
expression (\ref{zerodsolution})  
multiplied by the factor $\pi/b$ and with the parameter $\alpha$ replaced by 
$a+ib$.
How might we $\delta$-modify the ``action" $S$? The part $S_0$ should have some 
resemblance
to  $S$ and so a natural choice is
\begin{eqnarray}
S_{\delta}=&\Omega& \left(\Phi_1^{\dag}\Phi_1 -\Phi_2^{\dag}\Phi_2\right) +ib
\left(\Phi_1^{\dag}\Phi_1 +\Phi_2^{\dag}\Phi_2\right)\nonumber\\
&+&\delta\left[(a-\Omega)\left(\Phi_1^{\dag}\Phi_1 -\Phi_2^{\dag}\Phi_2\right) 
+\frac{i\lambda}{2}\left(\Phi_1^{\dag}\Phi_1 
-\Phi_2^{\dag}\Phi_2\right)^2\right],
\label{deltazerodaction}
\end{eqnarray}
where $\Omega=\Omega_1+i\Omega_2$, with $\Omega_1>0$ and $\Omega_2>b>0$.
Replacing $S$ by $S_{\delta}$ in Eq.\ (\ref{zerodcond}) and expanding in 
$\delta$ we find, however,
that the integrals are infinite since the integrands diverge for $s_1s_1^* - 
s_2s_2^*\rightarrow
-\infty$. The solution to this problem is to split the range of integration in 
Eq.\ (\ref{zerodcond})
 into the two regions
$s_1s_1^* > s_2s_2^*$ and $s_1s_1^* < s_2s_2^*$ and apply the LDE method 
separately to each integral
with $\Omega$  replaced by $\Omega^*=\Omega_1-i\Omega_2$  in the integral over 
the latter region. 
Since the   
two integrals
are complex conjugates of each other, it is sufficient to consider only one of 
them.
We are therefore led to 
consider the following $\delta$-modification:
\begin{equation}
I_{\delta}=2\ {\rm Re}\int d\chi_1^* d\chi_1 d\chi_2^* d\chi_2 ds_1^* ds_1 
ds_2^* ds_2\ 
s_1 s_1^* s_2 s_2^*\ \Theta(\Phi_1^{\dag}\Phi_1 -\Phi_2^{\dag}\Phi_2)\ 
\exp(iS_{\delta}),
\label{deltazerodcond}
\end{equation}   
where $S_{\delta}$ is given by Eq.\ (\ref{deltazerodaction}) and 
\begin{eqnarray}
\Theta(\Phi_1^{\dag}\Phi_1 -\Phi_2^{\dag}\Phi_2)&=&\frac{1}{2\pi 
i}\int_{-\infty}^{+\infty}
dq (q-i\epsilon)^{-1}\exp[iq (\Phi_1^{\dag}\Phi_1 
-\Phi_2^{\dag}\Phi_2)]\nonumber\\
&=&\Theta(s_1^* s_1 -s_2^* s_2) +(\chi_1^* \chi_1 -\chi_2^* \chi_2)\ 
\delta(s_1^* s_1 -s_2^* s_2 )
\nonumber\\
& &-\chi_1^* \chi_1 \chi_2^* \chi_2\ \delta'(s_1^* s_1 -s_2^* s_2).
\label{superstep}
\end{eqnarray}
The function $\Theta(\Phi_1^{\dag}\Phi_1 -\Phi_2^{\dag}\Phi_2)$ limits the 
integration range
to $s_1s_1^* > s_2s_2^*$, while preserving supersymmetry. We call this the 
supersymmetric step function.
Using the identity $\Theta(\Phi_1^{\dag}\Phi_1 -\Phi_2^{\dag}\Phi_2)+
\Theta(\Phi_2^{\dag}\Phi_2 -\Phi_1^{\dag}\Phi_1)=1$, we can check  that 
$I_{\delta=1}=I$. 
Expanding Eq.\ (\ref{deltazerodcond}) with respect to $\delta$ and then carrying 
out the integrals, we
obtain an expression which coincides with twice the real part of  expression 
(\ref{deltaexpand})  
multiplied by the factor $\pi/b$ and with the parameter $\alpha$ replaced by 
$a+ib$. 
Thus, we can immediately conclude that applying  the PMS condition 
and then taking the real part yields
a series which converges to the exact solution (\ref{zerodcondsolution}).  

Having found a way in which to apply the LDE method to the zero-dimensional 
conductance model, we can now  
try to apply the  method in the same way to the $S^{+-}$ and $S^{-+}$ terms in  
Eq.\ (\ref{conductance}).
The $\delta$-modified  actions $S_{\delta}^{+-}$ and $S_{\delta}^{-+}$ analogous 
to 
Eq.\ (\ref{deltazerodaction})  are clear.
As for the supersymmetric step function, the natural choice  is
\begin{equation}
\Theta\left(\int d{\bf r}\ \eta^{ab}_{mn}\Phi_m^{\dag}\Phi_n \right)
=\frac{1}{2\pi i}\int_{-\infty}^{+\infty}
dq (q-i\epsilon)^{-1}\exp\left(iq \int d{\bf r}\  
\eta^{ab}_{mn}\Phi_m^{\dag}\Phi_n\right).
\label{dsuperstep}
\end{equation}
The delta-modifications of the terms involving $S^{+-}$ and $S^{-+}$ in Eq.\ 
(\ref{conductance})
 are then just the analogues of Eq.\ (\ref{deltazerodcond}).           
We are faced with some difficulty, however, in  evaluating the terms in  the 
$\delta$-expansion series.
The difficulty lies in performing the  path integrals in the presence of
the step function (\ref{dsuperstep}). The first thought is to carry
out the path integrals first and then the $q$-integral appearing in the 
definition of the step function.
But interchanging the order of integration in this way makes the path integrals 
ill-defined for the same 
reason  as was
mentioned above for the ordinary integrals of the zero-dimensional model. 
One possible  resolution is to have a series expansion of 
the step function involving
only  polynomial and Gaussian functions of the superfields --- so that the path 
integrals
can be performed --- and to link the order of this expansion with
the order of the $\delta$-expansion. Relevant discussions can be found in Ref.\ 
\onlinecite{pernice}.  
  
\section{conclusion}

We have applied the LDE approximation  method to  the quantum dynamics 
of a single electron in a  random potential.
The LDE  method was originally developed  for the study of quantum field systems 
such as $\lambda\phi^4$ field theory. 
The idea that the LDE 
method might be applied to disordered systems as well comes from the resemblance 
of 
the path integral expressions of the   $\lambda\phi^4$ field theory $n$-point 
functions to the       
supersymmetric path integral expressions of quantities such as the averaged  
density of states and conductance for  Gaussian delta-function-correlated 
disorder. The supersymmetric
actions associated with these quantities contain several unusual features, such 
as Faddeev-Popov rather 
than Dirac kinetic term for the anticommuting fields, imaginary coupling, and 
Minkowskian field coordinate 
metric. We showed how the LDE method can be applied to such systems using as 
illustrative examples
the ground state energy of a supersymmetric anharmonic oscillator, 
a zero-dimensional density of states model, the  one-dimensional density of
states, and a zero-dimensional conductance model.

The next stage is to apply the LDE approximation to the density of states in 
$d\geq 2$ 
for Gaussian delta-function as well as other types of correlated  disorder. The 
calculations should
be similar to those for the $d=1$ density of states, the only essential new 
feature being the need to
regularize and renormalize. The ideas presented concerning the application of 
the LDE method to the 
conductance must also be developed further. 

We now finish with a brief description of another  possible application of the 
LDE method. 
This concerns the use of a supersymmetric non-linear $\sigma$-model 
to determine the conductivity critical exponent for the Anderson metal-insulator 
transition (see, e.g.,
Sects. 3 and 4 of Ref.\ \onlinecite{efetov} and also Ref.\ \onlinecite{kramer} 
for a review of the 
Anderson transition). 
The  $\sigma$-model is usually quantized using the $2+\epsilon$
expansion.\cite{hikami,wegner} However, the conductivity exponent for 
$\epsilon=1$ $(d=3)$ 
was found not to
agree with the accepted  value from numerical calculations.\cite{mackinnon} One 
possible 
reason put forward for this disagreement is the omission from the $\sigma$-model 
action of 
high-order gradient terms which may be relevant to the fixed point and 
$\epsilon$-expansion 
(see, e.g., Ref.\ \onlinecite{altshuler} for a review). 
Another possibility is that the $\sigma$-model is adequate,
but the method of perturbative quantization is not. With respect to the latter 
possibility, 
it would be of interest to try to apply the LDE or some related method to the 
supersymmetric
non-linear $\sigma$-model in order to  determine the conductivity exponent.

\acknowledgments

We would  like to thank  M. Chu, E. Hofstetter, M. Itoh,  H. F. Jones,
 A. MacKinnon, and L-H. Tang
for very helpful and stimulating discussions. We are especially grateful to H. 
F. Jones
for his close interest in the project and for suggesting improvements to the 
manuscript.    
Financial support from the EPSRC is also acknowledged.

\end{document}